\documentclass[conference]{IEEEtran}
\IEEEoverridecommandlockouts
\usepackage{booktabs}
\usepackage{graphicx}
\usepackage{amsfonts}
\usepackage{amsmath}
\usepackage{amssymb}
\usepackage{algorithm}
\usepackage{algorithmicx}
\usepackage{xcolor}
\usepackage{subfigure}
\usepackage{soul}

\usepackage[noend]{algpseudocode}
\usepackage{algpseudocode}

\begin{document}
\title{Spatial Interpolation-based Learned Index for Range and $k$NN Queries}

\author{\IEEEauthorblockN{Songnian Zhang}
\IEEEauthorblockA{\textit{Faculty of Computer Science} \\
\textit{University of New Brunswick}\\
Fredericton, NB, Canada \\
szhang17@unb.ca}
\and
\IEEEauthorblockN{Suprio Ray}
\IEEEauthorblockA{\textit{Faculty of Computer Science} \\
\textit{University of New Brunswick}\\
Fredericton, NB, Canada \\
sray@unb.ca}
\and
\IEEEauthorblockN{Rongxing Lu}
\IEEEauthorblockA{\textit{Faculty of Computer Science} \\
\textit{University of New Brunswick}\\
Fredericton, NB, Canada \\
rlu1@unb.ca}
\and
\IEEEauthorblockN{Yandong Zheng}
\IEEEauthorblockA{\textit{Faculty of Computer Science} \\
\textit{University of New Brunswick}\\
Fredericton, NB, Canada \\
yzheng8@unb.ca}
}

\maketitle

\renewcommand{\algorithmicrequire}{\textbf{Input:}}
\renewcommand{\algorithmicensure}{\textbf{Output:}}
\algrenewcommand\textproc{}
\renewcommand{\arraystretch}{1.5}

\begin{abstract}
A corpus of recent work has  revealed that the learned index can improve query performance while reducing the storage overhead. It potentially offers an opportunity to address the spatial query processing challenges caused by the surge in location-based services. Although several learned indexes have been proposed to process spatial data, the main idea behind these approaches is to utilize the existing one-dimensional learned models, which requires either converting the spatial data into one-dimensional data or applying the learned model on individual dimensions separately. As a result, these approaches cannot fully utilize or take advantage of the information regarding the spatial distribution of the original spatial data. To this end, in this paper, we  exploit it  by using the spatial (multi-dimensional) interpolation function as the learned model, which  can be directly employed on the spatial data.  Specifically, we design an efficient \underline{SP}atial inte\underline{R}polation funct\underline{I}on based \underline{G}rid index (SPRIG) to process the range and $k$NN queries. Detailed experiments are conducted on real-world datasets, and the  results  indicate that our proposed learned index can  significantly improve the performance in comparison with  the traditional spatial indexes and a state-of-the-art multi-dimensional learned index.
\end{abstract}


\section{Introduction}
\label{sec:intro}
As location-based services (LBS) have been widely deployed and have become highly popular,  spatial query processing  has attracted considerable interests in the research community.  Although several spatial indexes, such as R-tree and k-d tree, have been proposed to facilitate spatial query performance,  it is still challenging to process the spatial queries efficiently due to the rapidly growing volume of spatial data.  Recently, Kraska et al.~\cite{KraskaBCDP18} suggested substituting the traditional indexes with machine learned based indexes (also called \textit{learned index}).  Since then, several follow-up research projects~\cite{TangWDHWWC20, GalakatosMBFK19, FerraginaV20, QiLJK20, NathanDAK20} have shown that the learned index can indeed improve query performance by learned data distribution and query workload patterns.

Typically, there are two main aspects involving a learned index, namely,  a learned model and a local search.  The former is trained and used to quickly locate the approximate position of a search key, while the latter is responsible for refining the accurate position. Since the latter can be achieved by performing a local binary or exponential search, it is a fundamental but challenging topic to find a reasonable learned model and further employ it as the learned index. Existing learned indexes are constructed based on mainly one of two categories of learned models: machine learning~\cite{KraskaBCDP18} and  piecewise linear functions~\cite{GalakatosMBFK19}. However, to the best of our knowledge,  both of these learned models can be only applied in single dimensional data.   As a result, the current spatial learned indexes either transform multi-dimensional data into one-dimensional data before introducing the learned model as a foundation~\cite{WangFX019,DavitkovaM020} or apply a learned model on every single dimension~\cite{NathanDAK20}.  For this reason,  the question then arises,  ``\textit{Is there a learned model that can be directly applied to spatial (multi-dimensional) data and achieve better performance?}"

Aiming to address the above-mentioned question, in this paper,  we explore how to utilize spatial (two-dimensional) interpolation functions as the learned models to directly predict the position of a spatial search key. Based on this idea, we propose a \underline{SP}atial inte\underline{R}polation funct\underline{I}on based \underline{G}rid index (SPRIG) to support range and $k$NN queries over spatial data. In particular, we sample the spatial data to construct an adaptive grid and use  the sample data as inputs to fit a spatial interpolation function. Given a spatial search key, first, we can use the fitted spatial interpolation function to predict the approximate position of the key. Then, around the estimated position, we can conduct a local binary search to find the target key. However, it entails a new challenge: how to guarantee that the target key is in the local search range. To address this issue, we introduce the maximum estimation error based error guarantee, which is derived based on the query workload.  Furthermore, we propose efficient range and $k$NN query execution strategies using our proposed index. In these strategies, we take full advantage of the properties of the adaptive grid to facilitate the query executions, and a pivot based filtering technique is introduced to improve the $k$NN query performance. 

We conduct extensive experiments to evaluate our learned index, SPRIG. First, we evaluate five spatial interpolation functions and choose the \textit{bilinear interpolation} function, which has the best performance and estimation accuracy, as our learned model. Then, we compare SPRIG against the state-of-the-art multi-dimensional learned index Flood \cite{NathanDAK20}, along with a few spatial indexes. The experimental results involving real-world datasets show that: 1) SPRIG outperforms the alternative spatial indexes on range queries and is competitive on $k$NN queries in terms of execution time. In the best case, SPRIG is 3$\times$ faster than Flood with range queries and 9$\times$ faster than Flood with $k$NN queries; 2) SPRIG consumes less storage to achieve a favorable execution performance compared with the traditional indexes.  Our evaluations demonstrate that SPRIG can reduce the storage footprint of traditional spatial indexes by orders of magnitude.

The remainder of this paper is organized as follows. In Section~\ref{sec:rw}, we discuss the related work. Then, we introduce the spatial interpolation function, error guarantee, and pivot based filtering in Section~\ref{sec:bg}. After that, we present our SPRIG in Section~\ref{sec:index}, followed by performance evaluation in Section~\ref{sec:eva}. Finally, we draw our conclusion in Section~\ref{sec:con}.

\section{Related Work}
\label{sec:rw}
Kraska et al.~\cite{KraskaBCDP18} presented the idea of the learned index, which is based on learning the relationship between keys and their positions in a sorted array. They adopted a machine learning based technique as the learned model and built a recursive model index (RMI), which predicts the position of a search key within a known error bound.   Since then,  a variety of learned indexes was proposed to handle one-dimensional data. Recently,  Tang et al.~\cite{TangWDHWWC20} proposed a scalable learned index \textit{XIndex} based on RMI, which focuses on handling concurrent writes without affecting the query performance. Very differently, Galakatos et al.~\cite{GalakatosMBFK19} exploited the piecewise linear function as the learned model to build a data-aware index \textit{FITing-tree} that replaces leaf nodes of B+-tree with the learned piecewise linear functions. Unlike \textit{FITing-tree},  Ferragina  et al.~\cite{FerraginaV20} introduced a pure learned index \textit{PGM-index}  that does not mix the traditional data structure and learned model. However, their work still focuses on one-dimensional data and uses the existing linear learned model.

Naturally, the idea of the learned index has been extended to spatial and multi-dimensional data. Wang et al.~\cite{WangFX019} proposed a learned index \textit{ZM-index} for spatial queries. In that work, the authors utilized the Z-order curve to convert two-dimensional data into one-dimensional values, and then applied a machine learning model to predict a key's position on one-dimensional data.  Qi et al.~\cite{QiLJK20} refined the idea of \textit{ZM-index} and built a recursive spatial model index (RSMI). Before applying Z-order curve,  their work adopts a rank space-based transformation technique to mitigate the uneven-gap problem.  \textit{LISA}~\cite{Li0ZY020} is a disk-based spatial learned index that achieves low storage consumption and I/O cost.  In this work, the authors used a mapping function to map spatial keys into one-dimensional values and a monotone shard prediction function, which is similar to the piecewise linear functions, to predict the shard id for a given mapped value. Extending to multi-dimensional data, \textit{ML-index} ~\cite{DavitkovaM020} is an RMI based learned index. It first converts the multi-dimensional data into one dimension by employing the i-Distance technique. Based on the one-dimensional data, \textit{ML-index} uses the RMI to estimate the approximate position of a search key.  Recently proposed index Flood~\cite{NathanDAK20} can also support multi-dimensional data and is very relevant to our work. It  adopts the \textit{FITing-tree} as the building block to predict the key's position on a single dimension. By integrating $d -1$ dimensions' positions, where $d$ is the number of dimensions,  Flood can locate the cell that covers the search key.

\section{Background}
Before delving into the details of SPRIG, in this section, we introduce three basic concepts: 1) Spatial Interpolation Function; 2) Error Guarantee; and 3) Pivot Based Filtering, which serve as the building blocks of the proposed index.

\begin{figure}[!t]
%\vspace{-0.2cm}
\centering
\includegraphics[width=\linewidth]{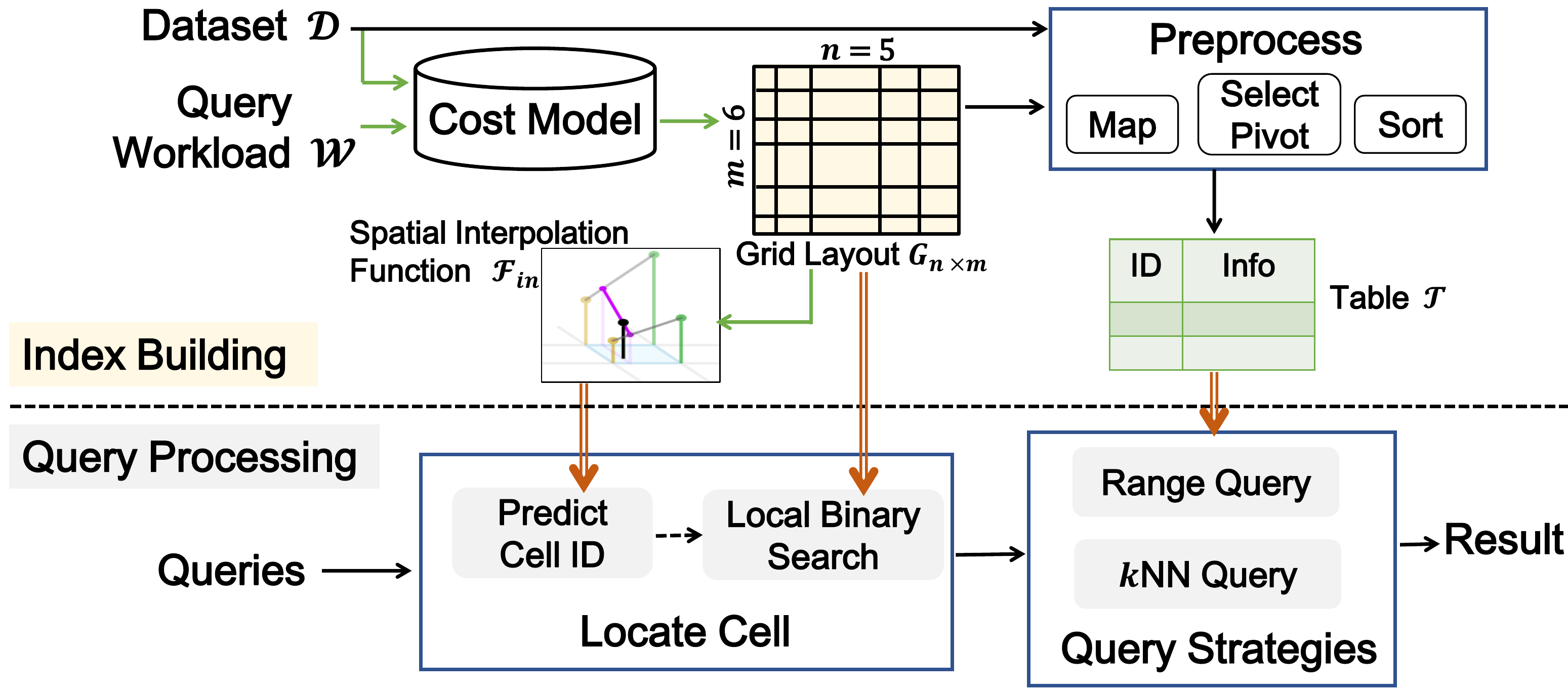}
\vspace{-0.64cm}
\caption{System Architecture of SPRIG}
\label{fig.scheme.sys}
\vspace{-0.3cm}
\end{figure}

\label{sec:bg}
\subsection{Spatial Interpolation Function}
Given a set of 2-dimensional sample points $\{(x_i, y_i) |  1 \le i \le n\}$ and their corresponding values $\{v_i =f(x_i, y_i) |1 \le i \le n\}$,  one can construct a spatial (two-dimensional) interpolation function $\mathsf{F}_{in} = f(x, y)$ that passes through all these sample points~\cite{MitasM99}.  Afterward, given any point $(x,y)$, it is easy to estimate the value of  $f(x, y)$ with the interpolation function. Borrowing the idea from the learned index~\cite{KraskaBCDP18}, if we treat $v_i$ as the position of the point $(x_i, y_i) $,  we can use  $\mathsf{F}_{in}$ to quickly estimate the position of any given point. Moreover, if the sample points are not random but can represent the distribution of the original spatial dataset,  we can use fewer sample points to fit the spatial interpolation function for estimating positions. It indicates that the spatial interpolation function can learn the spatial position distribution with a lower storage overhead, which fits well with the goal of the learned index. Therefore, it is feasible and promising to exploit  the spatial interpolation function $\mathsf{F}_{in}$ as the learned model.

\subsection{Error Guarantee}
For a learned index,  it is essential to conduct a local search after predicting the position of a given point. Generally, the local search range is $[pos -\textit{eg}, ~pos+\textit{eg}]$, where $pos$ is the predicted position and $\textit{eg}$ is the estimation error, also called error guarantee. Consequently,  $\textit{eg}$ is an essential concept in a learned index.  Different from ~\textit{FITing-tree}~\cite{GalakatosMBFK19} (used in Flood~\cite{NathanDAK20}), we adopt the \textit{maximum estimation error} as the error guarantee in our scheme, which can be determined by a query workload $\mathcal{W}$.
\begin{equation}
\label{equ.eg}
 \textit{eg} = max(\mathsf{F}_{in}(q_x, q_y) - f(q_x, q_y)),
 \end{equation}
where $(q_x, q_y) \in \mathcal{W}$. Regarding a spatial point,  its spatial position can be determined by its $x$ and $y$ coordinates. Therefore, in our scheme, we project the estimated spatial position to $x$ dimension  and $y$ dimension and obtain two error guarantees, i.e,
\[
 \begin{split}
 \textit{eg}_x = max(\mathcal{P}_x(\mathsf{F}_{in}(q_x, q_y)) - \mathcal{P}_x(f(q_x, q_y))); \\
 \textit{eg}_y = max(\mathcal{P}_y(\mathsf{F}_{in}(q_x, q_y)) - \mathcal{P}_y(f(q_x, q_y))),
\end{split}
\]
where $\mathcal{P}_x$() and $\mathcal{P}_y()$ project a spatial position to $x$ dimension and $y$ dimension, respectively.

\subsection{Pivot Based Filtering}
Assume that a set $\mathcal{D}$ contains $n$ 2-dimensional points, i.e, $\mathcal{D}=\{p_i =(x_i, y_i) |  1 \le i \le n\}$. Also,  we define a distance based range query $\mathsf{Q}_c=\{q_c, r\}$, where $q_c$ is a point,  and $r$ is a radius. Launching a query $\mathsf{Q}_c$ over $\mathcal{D}$ means  one would like to find points in $ \mathcal{D}$ that satisfy $d(p_i, q_c) \le r$, where $d(\cdot)$ calculates the Euclidean distance between two spatial points. The intuitive solution is to scan and check points in $\mathcal{D}$ one by one. However, it is inefficient. Although we can build tree based index structures, such as k-d tree and M-tree~\cite{CiacciaPZ97}, to speed up the query processing, it will incur much extra storage.  In order to improve the performance of  $\mathsf{Q}_c$ without introducing too much storage overhead, we adopt a pivot based filtering technique. The key idea is to select a virtual pivot $p_v$ for $\mathcal{D}$ and calculate distances $d(p_i, p_v)$ for each point $p_i$. Then, sort $\mathcal{D}$ according to the corresponding $d(p_i, p_v)$. When performing a query $\mathsf{Q}_c$ over $\mathcal{D}$, we can only calculate the distance $d(q_c, p_v)$ and check the points whose distances $d(p_i, p_v)$ lie in $[d(q_c, p_v) -r,  d(q_c, p_v) + r]$, instead of all points in  $\mathcal{D}$. This optimization technique is based on triangle inequality, and its correctness is as follows:
\begin{align*}
& |d(p_i, p_v) - d(q_c, p_v)| \le d(p_i, q_c) \le r \\
& \Rightarrow  -r \le d(p_i, p_v) - d(q_c, p_v) \le r \\
& \Rightarrow  d(q_c, p_v) -r \le d(p_i, p_v)  \le d(q_c, p_v) + r.
\end{align*}

The above inequality indicates that if $p_i$ falls within $\mathsf{Q}_c$, it must satisfy $d(p_i, p_v) \in [d(q_c, p_v) -r,  d(q_c, p_v) + r]$. Thus, we can narrow down the scan range from the whole dataset $\mathcal{D}$ to the points that have distances  $d(p_i, p_v)$ in $[d(q_c, p_v) -r,  d(q_c, p_v) + r]$.

%Figures query
\begin{figure*}[!t]
%\vspace{-0.8cm}
\centering
\setlength{\abovecaptionskip}{0.cm}
\subfigcapskip=-10pt
\subfigure[Range Query]{
    \begin{minipage}[t]{0.43\textwidth}
    \centering
    \includegraphics[height=0.48\textwidth,width=\textwidth]{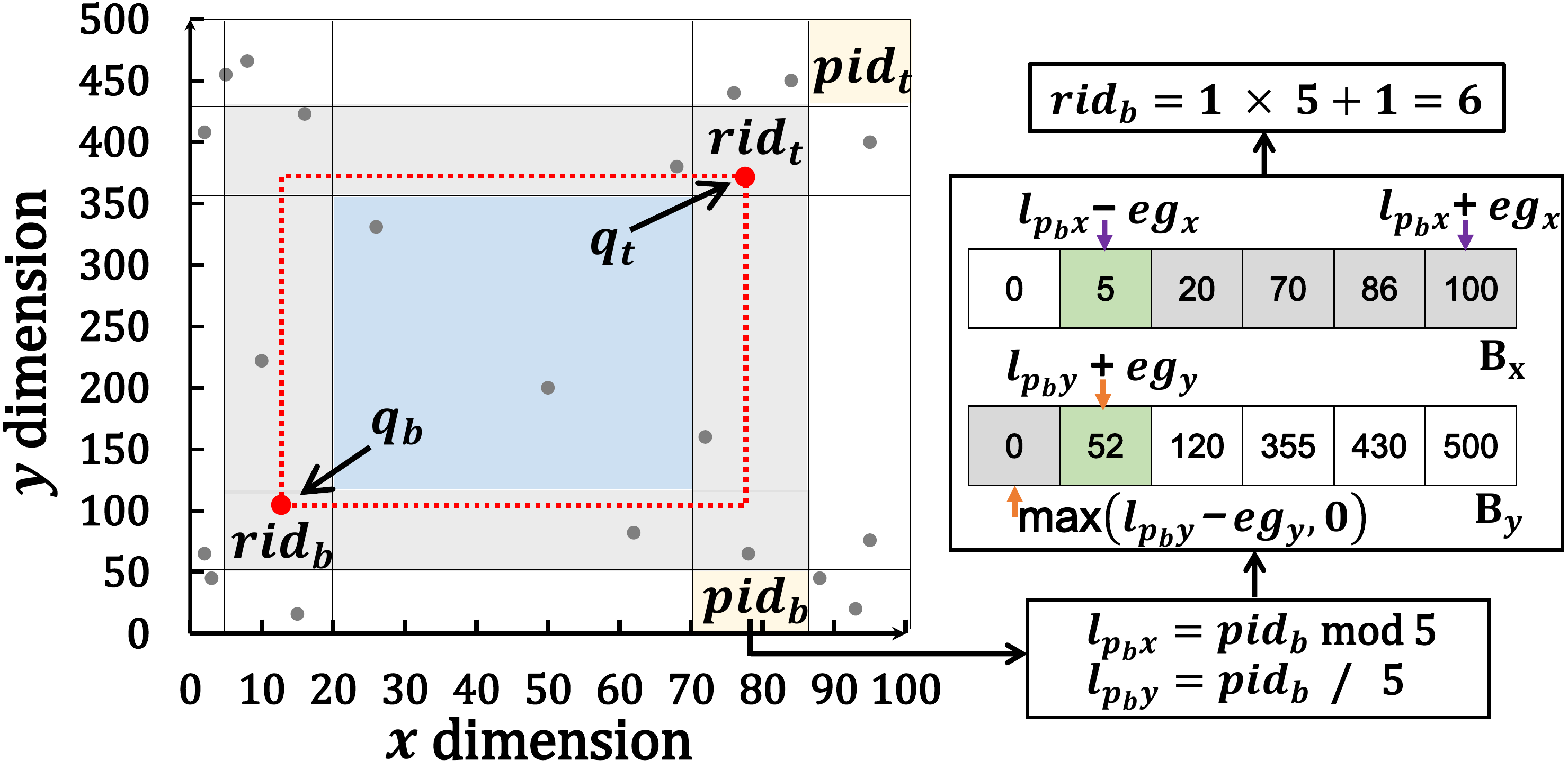}
    \label{fig.scheme.rq}
    \end{minipage}%
}\hspace{10mm}%
\subfigure[$k$NN Query]{
    \begin{minipage}[t]{0.43\textwidth}
    \centering
    \includegraphics[height=0.48\textwidth,width=\textwidth]{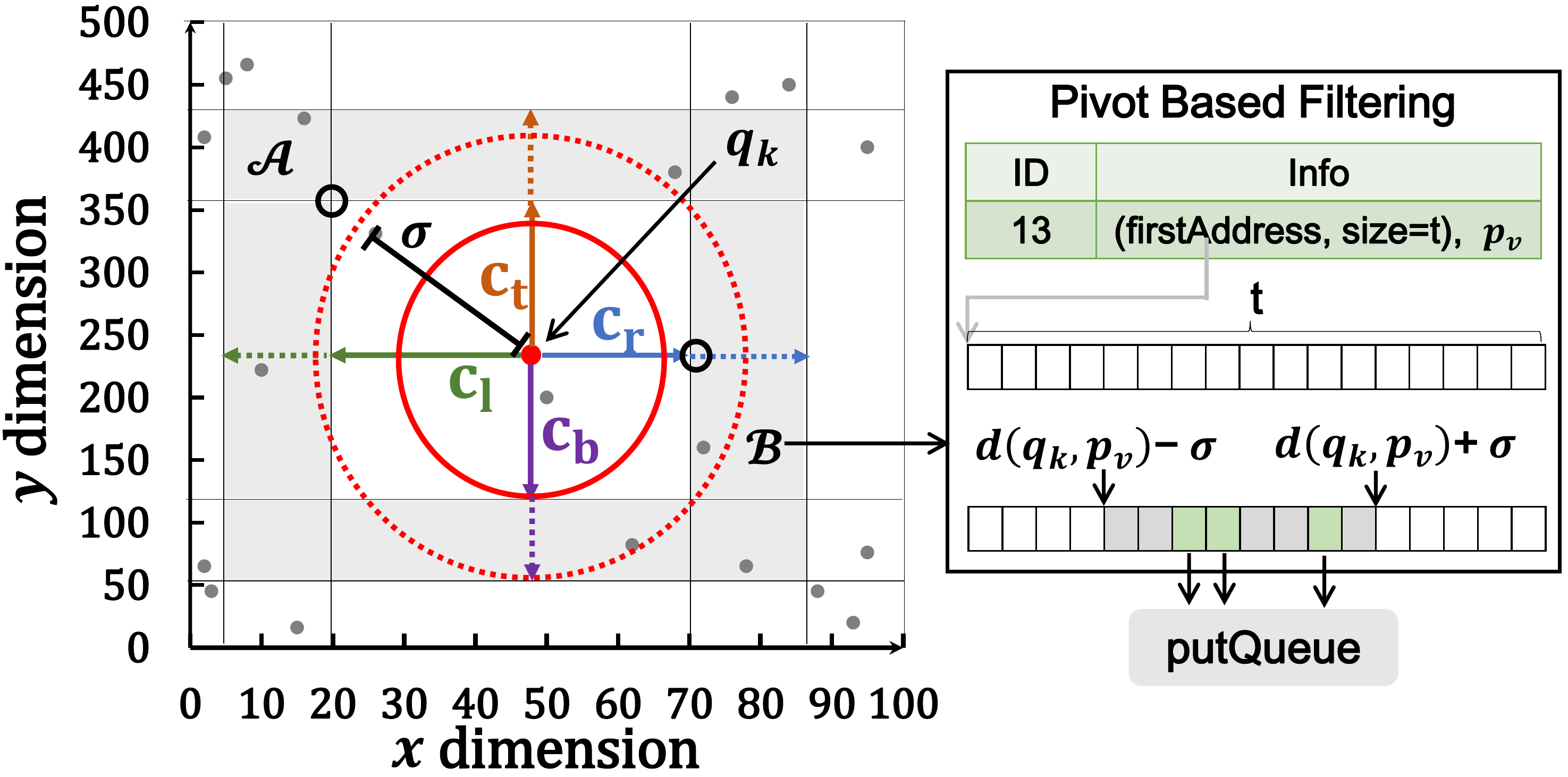}
    \label{fig.scheme.knn}
    \end{minipage}%
}%
%\vspace{-0.1cm}
\caption{Processing queries on SPRIG. A 5 $\times$ 5 grid layout. (a) Range Query. The red dashed rectangle indicates the range query window, and the yellow cells are estimated with $\mathsf{F}_{in}$. Assume that ($\textit{eg}_x$=2, $\textit{eg}_y$=1), we take finding $\textit{rid}_b$ as an example.  (b) $k$NN Query. Spread the search range to the outer-layer adjacent cells. The hollow circles represent the closest point of cell $\mathcal{A}$ and cell $\mathcal{B}$ to $q_k$. Assume that cell  $\mathcal{B}$ has $t$ records, we use the pivot based filtering to process the records in the cell.}
\label{fig.scheme.query}
\vspace{-0.2cm}
\end{figure*}

%$\mathsf{B_x}$=$\{0, 5,20,70,86,100\}$, $\mathsf{B_y}$=$\{0,52, 120, 355,430, 500\}$

\section{Our proposed index-SPRIG}
\label{sec:index}
In this section,  we present the details of our proposed index, SPRIG. Fig.~\ref{fig.scheme.sys} depicts the system architecture of our index, which is comprised of two parts: index building and query processing. In the following, we first discuss how to build the learned index. Then, we describe the detailed query processing on our index.

\subsection{Index Building}
\label{sub.index.build}
Our SPRIG mainly consists of three components: 1) An $n \times m$ grid  layout $\mathsf{G}_{n \times m}$, where $n$ is the number of columns along $x$ dimension and $m$ is for $y$ dimension; 2) A table $\mathcal{T}$; and 3) A spatial interpolation function $\mathsf{F}_{in}$ based learned model, as shown in Fig.~\ref{fig.scheme.sys}. Here, we may use some parameters for our index, which will be further discussed in Section~\ref{sub.cost.model}. To build the $n \times m$ grid  layout, we first  find $n-1$ boundaries on $x$ dimension to generate $n$ non-equal size columns, and add these boundaries into a set $\mathsf{B_x}$. Our goal is to make data records evenly distributed across columns, i.e., each column has a roughly equal number of records (in this paper, we use ``point'' and ``record'' interchangeably). For $y$ dimension, there will be $m-1$ boundaries in set $\mathsf{B_y}$.  We denote the maximum and minimum values in $x$ dimension as $\{x_\textit{min}, x_\textit{max}\}$, while those are $\{y_\textit{min}, y_\textit{max}\}$ for $y$ dimension.  After adding  $\{x_\textit{min}, x_\textit{max}\}$ into $\mathsf{B_x}$ and $\{y_\textit{min}, y_\textit{max}\}$ into $\mathsf{B_y}$,  we separately sort $\mathsf{B_x}$ and  $\mathsf{B_y}$ in an increasing order, where $|\mathsf{B_x}|=n+1$ and  $|\mathsf{B_y}|=m+1$.  Totally, there are  $n \times m$ cells for the grid, and we have $\mathsf{G}_{n \times m} =(\mathsf{B_x}, \mathsf{B_y})$. Algorithm~\ref{algo.scheme.bg} shows the process of building the grid. 

Next, we allocate integers in the range [0, $n \times m-1$] as cell ids along $x$ dimension and define a 2-dimensional array $\mathsf{C_{id}}$ to index these cell ids:  $\mathsf{C_{id}}[i][j] = j \cdot n + i,  0 \le i < n~$ and $~  0 \le j < m$. Afterward, we can build a table  $\mathcal{T}$ to map the cell id to the covered records, in which the key is the cell id and the value is a pair $(\textit{firstAddress}, \textit{size})$ indicating the pointer to the first record and the number of records in the cell. Based on  $\mathsf{G}_{n \times m} $ and $\mathsf{C_{id}}$, we can fit a spatial interpolation function  $\mathsf{F}_{in}$.   In particular,  we treat $\{\mathsf{B_x}, \mathsf{B_y}\}$ as inputs and $\mathsf{C_{id}}$ as the desired estimation values to determine $\mathsf{F}_{in}$, i.e., $\mathsf{C_{id}} \gets \mathsf{F}_{in}(\mathsf{B_x}, \mathsf{B_y})$.  Besides, to speed up the distance related search, for example, $k$NN queries, we employ the pivot based filtering technique, which drives us to select a pivot $p_v$ in a cell and sort the records (in a cell) according to the distances $d(p_i, p_v)$.  In this paper, we adopt the $k$-means clustering to select pivots for each cell.  It is worth noting this technique does not incur much extra storage except for pivot points.

\subsection{Query Processing}
\label{sub.index.query}
In this paper, we focus on the range and $k$NN queries, in which the point query is implicitly included. Next, we describe our approaches.

\textbf{Range Query}.
For spatial data, the range query identifies the records ${p}_{i}$  that fall within $\mathsf{Q}_r$=$(q_b, q_t)$, where $q_b=[b_x, b_y]$ and $q_t=[t_x, t_y]$ are the bottom left and top right point of a query window, respectively.   The main idea of processing a range query is to convert it to locate the cell ids of $q_b$ and $q_t$. Here, we take locating  $q_b$'s cell id as an example, and the steps are as follows.
 \begin{itemize}
   \item \textit{Step-1}.  Given  $q_b$, we can use  $\mathsf{F}_{in}$ to obtain a predicted cell id $\textit{pid}_b$.  It is simple to calculate the locations of $\textit{pid}_b$ in set $\mathsf{B_x}$ and $\mathsf{B_y}$. That is,  $l_{px} = \textit{pid}_b \mod n$ and $l_{py} = \textit{pid}_b~/~n$.
   \item \textit{Step-2}.  Given a pair of error guarantee $(\textit{eg}_x, \textit{eg}_y)$,  by applying local binary searches on  $\mathsf{B_x}$ and  $\mathsf{B_y}$, we can obtain the real $x$ and $y$ locations of the search point, denoted as $l_{rx}$ and $l_{ry}$. The search range on $\mathsf{B_x}$ set is $[l_{px}-\textit{eg}_x, ~l_{px}+\textit{eg}_x ]$, while it is  $[l_{py}-\textit{eg}_y, ~l_{py}+\textit{eg}_y ]$ on  $\mathsf{B_y}$ set.
   \item \textit{Step-3}. Finally, we can obtain the real cell id of $q_b$, namely,  $\textit{rid}_b = l_{ry} \cdot n + l_{rx}$.
 \end{itemize}
Algorithm~\ref{algo.scheme.rid} formally outlines the above steps to obtain the real cell id of a search point. Similarly, we can get the real cell id  $\textit{rid}_t$ of $q_t$. After that, the  range query result can be collected with the following strategy:

1) For cells intersected by query window $\mathsf{Q}_r$, we scan the records in these cells and put the records that fall within $\mathsf{Q}_r$ into the range query result.

2) For cells contained inside query window $\mathsf{Q}_r$, we directly add all the records covered in these cells into  the result.

As shown in Fig.~\ref{fig.scheme.rq}, the intersected cells must be on the vertical and horizontal lines of $\textit{rid}_b$ and $\textit{rid}_t$. Therefore, it is easy to determine the intersected cells (gray area) and contained cells (blue area).
% Algorithm~\ref{algo.scheme.rq} formally outlines the entire range query process.

\setlength{\textfloatsep}{9pt}% Remove \textfloatsep
\setcounter{algorithm}{0}
\begin{algorithm}[!t]
\small
\caption{GetRealCellId}
\begin{algorithmic}[1]
    \Require
      A given point  $(x, y)$;  The grid layout,  $\mathsf{G}_{n \times m} =(\mathsf{B_x}, \mathsf{B_y})$; A predicted cell id of the given point,  \textit{pid};  Trained error guarantee $\textit{eg} = (\textit{eg}_x, \textit{eg}_y)$;
    \Ensure
      The real cell id   \textit{rid} of the given point $(x, y)$.
      \State  n $\gets |\mathsf{B_x}| -1$
      \State  $l_{px}$ $\gets$ ${pid} \mod$ n;  \hspace{3mm} $l_{py}$ $\gets$ ${pid}  ~/~$ n
      \State  $l_{rx}$ $\gets$ BinarySearch($\mathsf{B_x}$, $x$, $l_{px}-\textit{eg}_{x}$, $l_{px}+\textit{eg}_{x}$)
      \State  $l_{ry}$ $\gets$ BinarySearch($\mathsf{B_y}$, $y$, $l_{py}-\textit{eg}_{y}$, $l_{py}+\textit{eg}_{y}$)
      \State \Return $l_{ry} \cdot n + l_{rx}$
\end{algorithmic}
\label{algo.scheme.rid}
\end{algorithm}

% Algorithm Build Grid
\setlength{\textfloatsep}{8pt}% Remove \textfloatsep
\setcounter{algorithm}{1}
\begin{algorithm}[!t]
\small
\caption{Build Grid}
\begin{algorithmic}[1]
    \Require
      A spatial dataset, $\mathcal{D}$; The number of columns along $x$ dimension, $n$;  The number of columns along y dimension, $m$.
    \Ensure
      A grid layout $\mathsf{G}_{n \times m} =(\mathsf{B_x}, \mathsf{B_y})$.
     \State $\textit{mapX} \gets \varnothing$; ~~~  $\textit{mapY} \gets \varnothing$
    \For{$\textit{each entry in}~\mathcal{D}$}
        \State  \textit{cntX} $\gets$ ($\textit{mapX}.get(\textit{entry}.x) is~\varnothing$) ? 1:$\textit{mapX}.get(\textit{entry}.x) + 1$
        \State $\textit{mapX}.put(\textit{entry}.x, \textit{cntX})$

        \State  \textit{cntY} $\gets$ ($\textit{mapY}.get(\textit{entry}.y) is~\varnothing$) ? 1:$\textit{mapY}.get(\textit{entry}.y) + 1$
        \State $\textit{mapY}.put(\textit{entry}.y, \textit{cntY})$
    \EndFor
    \State $(x_\textit{min}, y_\textit{min}) \gets \textit{findMin}(\mathcal{D})$;~~$(x_\textit{max}, y_\textit{max}) \gets \textit{findMax}(\mathcal{D})$
    \State \textit{mapX}.sortByKey();~~~ ~~~\textit{mapY}.sortByKey()    ~~~~~\#Ascending
    \State \textit{avgX} $\gets \mathcal{D}.\textit{size}/\textit{n}$; ~~~\textit{avgY} $\gets\mathcal{D}.\textit{size}/\textit{m}$
     \State  $\mathsf{B_x} \gets \textit{getBoundary}(\textit{mapX},\textit{avgX}, x_\textit{min}, x_\textit{max})$
     \State  $\mathsf{B_y} \gets \textit{getBoundary}(\textit{mapY},\textit{avgY}, y_\textit{min}, y_\textit{max})$
     \State \Return{($\mathsf{B_x},  \mathsf{B_y}$)}
    \State
    \Function {$\textit{getBoundary}$}{$\textit{map}, \textit{avg}, \textit{min}, \textit{max}$}
    	\State $\mathsf{B} \gets \varnothing$;~~~$\textit{cnt} \gets 0$;~~~ $\textit{pre} \gets 0$
    	\State $\mathsf{B}.add(\textit{min})$
        \For{$\textit{each entry in map}$}
            \State $\textit{singleCnt} \gets \textit{entry.value}$
            \If{$\textit{singleCnt} > \textit{avg}$}
                \State $\mathsf{B}.add(\frac{\textit{entry.key} + \textit{pre}}{2})$
                \State $\textit{pre} \gets \textit{entry.key} $
                \State $\textit{cnt} \gets 0$
                \State $\textit{continue}$
            \EndIf
             \State $\textit{cnt} \gets \textit{cnt} + \textit{singleCnt}$
              \If{$\textit{cnt} > \textit{avg}$}
              	    \State $\mathsf{B}.add(\frac{\textit{entry.key} + \textit{pre}}{2})$
                    \State $\textit{cnt} \gets 0$
	      \Else
		    \State $\textit{pre} \gets \textit{entry.key} $
              \EndIf
        \EndFor
        \State $\mathsf{B}.add(\textit{max})$
        \State \Return {$\mathsf{B}$}
    \EndFunction
\end{algorithmic}
\label{algo.scheme.bg}
\end{algorithm}
% End

% Algorithm kNN Query
\setlength{\textfloatsep}{8pt}% Remove \textfloatsep
\setcounter{algorithm}{2}
\begin{algorithm}[!t]
\small
\caption{$k$NN Query}
\begin{algorithmic}[1]
    \Require
      A $k$NN query $Q_\textit{$k$NN} = (q_k, k)$;  Our index,  $\mathsf{G}_{n \times m} =(\mathsf{B_x}, \mathsf{B_y})$,  $\mathcal{T}$, and $\mathsf{F}_{in}$; Trained error guarantee $\textit{eg} = (\textit{eg}_x, \textit{eg}_y)$.
    \Ensure
      A priority queue $\textit{queue}$ contains the $k$ closest points.
     \State  $\textit{pid}$ $\gets$ $F_{in}(q_k.x, q_k.y)$;
    \State  ${rid}$ $\gets$ $\textit{getRealCellId}(q_k.x, q_k.y, \textit{pid}, \mathsf{B_x}, \mathsf{B_y}, \mathsf{eg})$
     \State  $l_{rx}$ $\gets$ ${rid} \mod n$; ~~$l_{ry}$ $\gets$ ${rid} / n$
     \State $e_c \gets 0$;   ~~~ $k_{cnt} \gets 0$;~~~ $queue \gets \varnothing$;
     \While{$k_{cnt} < k$}
     	 \State $c_b \gets q_k.y - \mathsf{B_y}[l_{ry}-e_c]$;~~~$c_t \gets \mathsf{B_y}[l_{ry}+1+e_c] -  q_k.y$
     	 \State $c_l \gets  q_k.x - \mathsf{B_x}[l_{rx}-e_c]$; ~~~$c_r \gets \mathsf{B_x}[l_{rx}+1+ e_c] -  q_k.x$
     	 \State $r \gets \textit{min}(c_b, c_t, c_l, c_r)$
     	 \If{$e_c > 0$}
     	 	\State $\textit{cells} \gets \textit{collectAdjacent}(q_k, \mathsf{G}_{n \times m}, \mathcal{T}, e_c)$
     	 	\For{$\textit{each cell in cells}$}
     	 		\If{$d(q_k, cell.p_c) < \textit{queue.peek.dist}$}
     	 			\State $\textit{records} \gets \textit{PivotFilter}(\textit{cell}, q_k, \textit{queue})$
     	 			\For {$\textit{each entry in records}$}
     	 				\State $\textit{putQueue}(\textit{entry}, \textit{queue}, q_k, k_{cnt}, r, k)$
				\EndFor
			\EndIf
		\EndFor
	\Else
	      \For {$\textit{each entry in} ~~\mathcal{T}.get(rid)$}
     	 		\State $\textit{putQueue}(\textit{entry}, \textit{queue}, q_k, k_{cnt},r, k)$
		\EndFor
	\EndIf
	\State $e_c \gets e_c + 1 $   	
     \EndWhile

    \State
    \Function {$\textit{putQueue}$}{$\textit{entry}, \textit{queue}, q_k, k_{cnt}, r, k$}
    	\State $\textit{dist} \gets \textit{getDisance}(q_k, \textit{entry})$
    	\If{$queue.size < k$}
    		\State $\textit{queue.offer(entry)}$
    		\If{$\textit{dist} \le r$}
    			\State $ k_{cnt} \gets  k_{cnt} + 1$
		\EndIf
	\ElsIf{$\textit{queue.peek().dist} > dist$}
		\State $\textit{queue.poll()}$
		\State $\textit{queue.offer(entry)}$
		\If{$\textit{dist} \le r$}
    			\State $ k_{cnt} \gets  k_{cnt} + 1$
		\EndIf
	\EndIf
    \EndFunction
\end{algorithmic}
\label{algo.scheme.kq}
\end{algorithm}
% End

\textbf{$k$NN Query}.
In this paper, we denote the $k$NN query as $Q_\textit{$k$NN} = (q_k, k)$, where $q_k$ is a point, and $k$ is the number of nearest neighbors. The formal definition is: $\forall p_i \in S,  \forall  p_j \in \mathcal{D} \setminus S, d(q_k, p_j) \ge d(q_k, p_i)$,   where $S$ is the result set of $k$NN query.  For $k$NN queries, our solution is to locate the real cell id of the query point  $q_k$. Then, starting from the cell, we recursively spread the search range and incrementally check the records in the outer-layer adjacent cells until the result is complete.  To facilitate improved  performance,  we employ two pruning techniques. One is the closest point pruning technique that is used to determine whether a cell should be checked. A cell may have different closest points to the different query points. However, in our scheme, since it is simple to obtain the locations $(l_{rx}, l_{ry})$ of the located real cell,   we can get the closest point of the cell, denoted as $p_c$, with the help of $\mathsf{B_x}$ and  $\mathsf{B_y}$ easily. In Fig.~\ref{fig.scheme.knn}, cell $\mathcal{A}$  has the closest point  $(\mathsf{B_x}[l_{rx}], \mathsf{B_y}[l_{ry} + 1]) $, and the cell $\mathcal{B}$'s closest point is $(\mathsf{B_x}[l_{rx} + 1], q_k.x)$. The other pruning technique is the pivot based filtering that can filter out records in a cell by the principle of the triangle inequality.  By employing the spreading outwards strategy,  closest-point pruning, and pivot based filtering, our $k$NN solution works as follows and is formally depicted in Algorithm~\ref{algo.scheme.kq}:
\begin{itemize}
   \item \textit{Step-1}.  Locate the real cell of $q_k$. With $\mathsf{F}_{in}$ and local binary search, we obtain a real cell id $\textit{rid}$ of  the query point $q_k$, which is shown in $\textit{line 1}$ and $\textit{line 2}$.
   \item \textit{Step-2}.  Calculate the distances from  $q_k$ to the borders of the located cell. We denote them as $c_t$, $c_b$, $c_l$, and $c_r$, as shown in Fig.~\ref{fig.scheme.knn}. Then, we obtain a radius $r$=$\textit{min}$($c_t$, $c_b$, $c_l$, $c_r$). See $\textit{line 6}$ - $\textit{line 8}$ for this step.
   \item \textit{Step-3}.  Scan records in the cell. After filtering the records by the pivot based filtering technique, we put a record $p_i$ into a priority queue if the queue's size is less than $k$ or $d(p_i, q_k) < \sigma$, where $\sigma$ is the distance between \textit{queue.peek} and $q_k$. Besides, we use a counter $k_{\textit{cnt}}$ to record the result size and increase it if $d(p_i, q_k) \le r$. In Algorithm~\ref{algo.scheme.kq}, we use a separate function $\textit{putQueue()}$ ($\textit{line 21}-\textit{line 31}$) to show the details of this step.
   \item \textit{Step-4}. If $k_{\textit{cnt}} < k$, we expand the search range and calculate a new radius $r$=$\textit{min}$($c_t$, $c_b$, $c_l$, $c_r$), where $c_t$, $c_b$, $c_l$, $c_r$ are the distances from $q_k$ to the borders of outer layer adjacent cells. For each adjacent cell, we  first check whether $d(p_c, q_k) > \sigma$. If yes, we skip the cell. Otherwise,  perform \textit{Step-3}. As shown in Fig.~\ref{fig.scheme.knn}, we would skip cell $\mathcal{A}$ and further process cell $\mathcal{B}$.
\end{itemize}

Repeat $\textit{Step-4}$ and $\textit{Step-3}$ until  $k_{\textit{cnt}} \ge k$. Eventually, the priority queue holds the result of  $k$NN query. Note that, since the \textit{collectAdjacent()} and  \textit{PivotFilter()} functions are straightforward, we do not present them in Algorithm~\ref{algo.scheme.kq} due to the limited space.

%Figures Range query
\begin{figure*}[!t]
%\vspace{-0.8cm}
\centering
\setlength{\abovecaptionskip}{0.cm}
\subfigcapskip=-10pt
\subfigure[Tweet200K]{
    \begin{minipage}[t]{0.22\textwidth}
    \centering
    \includegraphics[height=0.78\textwidth,width=\textwidth]{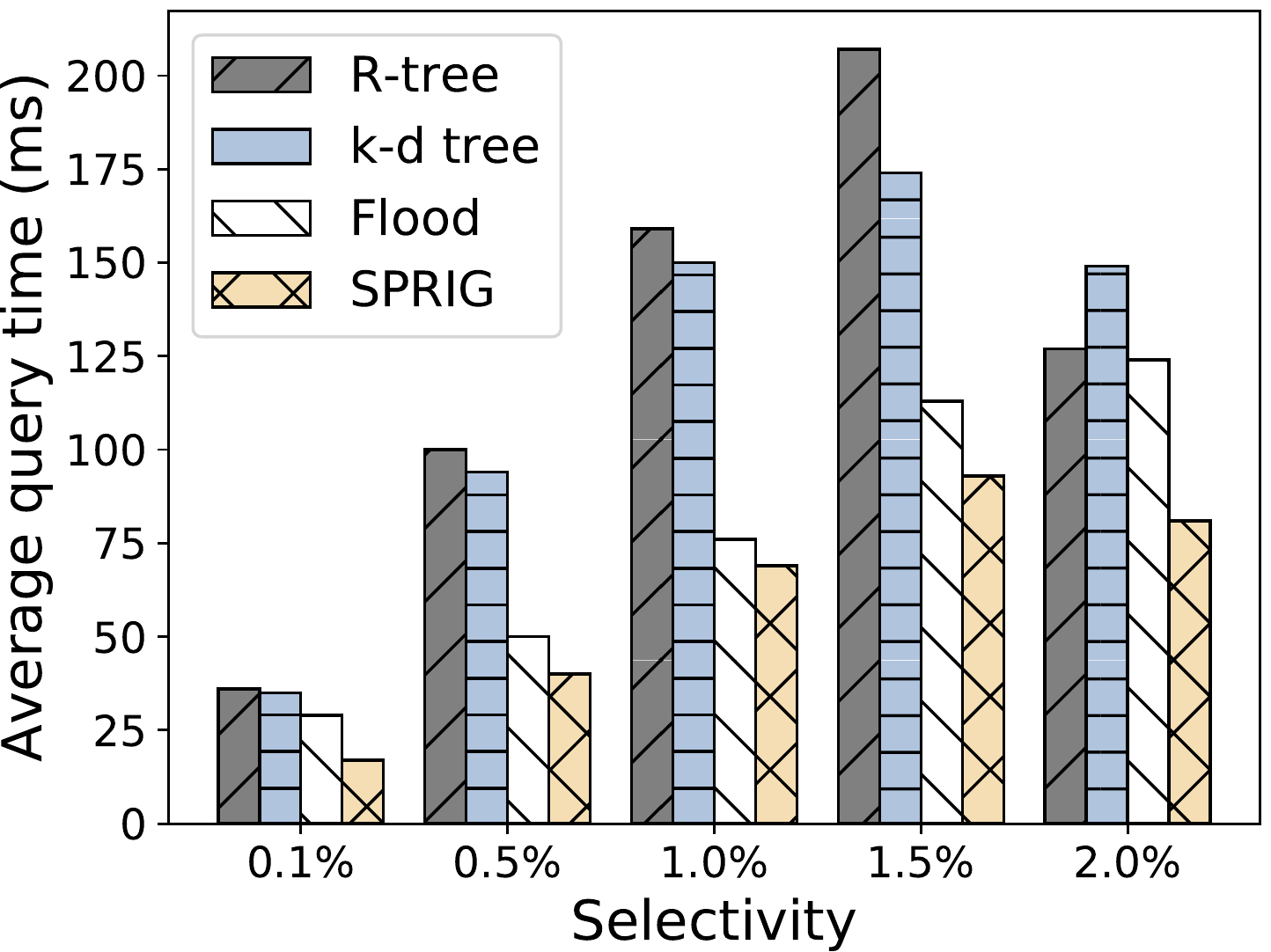}
    \label{fig.rq.200k}
    \end{minipage}%
}\hspace{10mm}%
\subfigure[Tweet2M]{
    \begin{minipage}[t]{0.22\textwidth}
    \centering
    \includegraphics[height=0.78\textwidth,width=\textwidth]{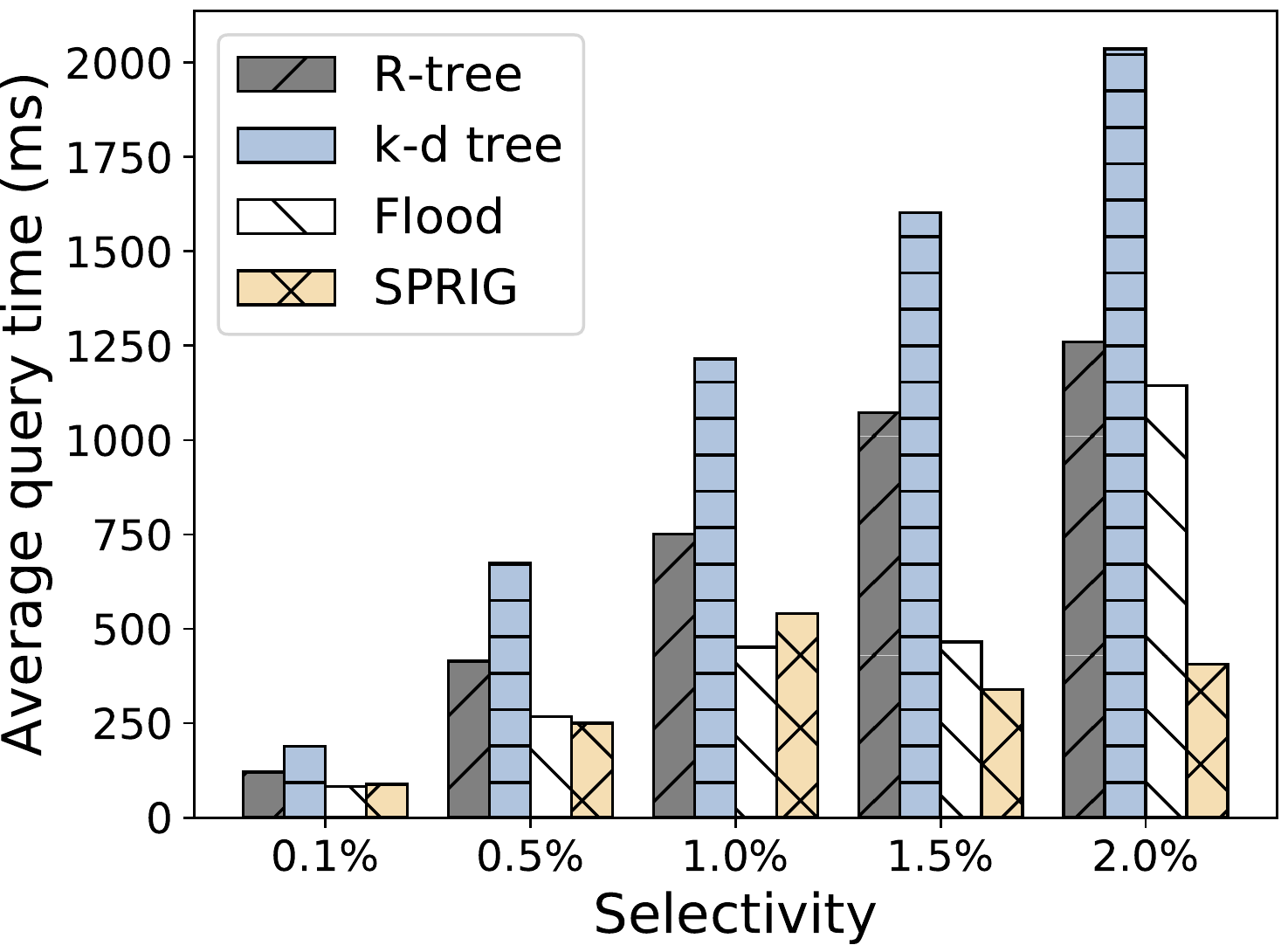}
    \label{fig.rq.2m}
    \end{minipage}%
}\hspace{10mm}%
\subfigure[Tweet20M]{
    \begin{minipage}[t]{0.22\textwidth}
    \centering
    \includegraphics[height=0.78\textwidth,width=\textwidth]{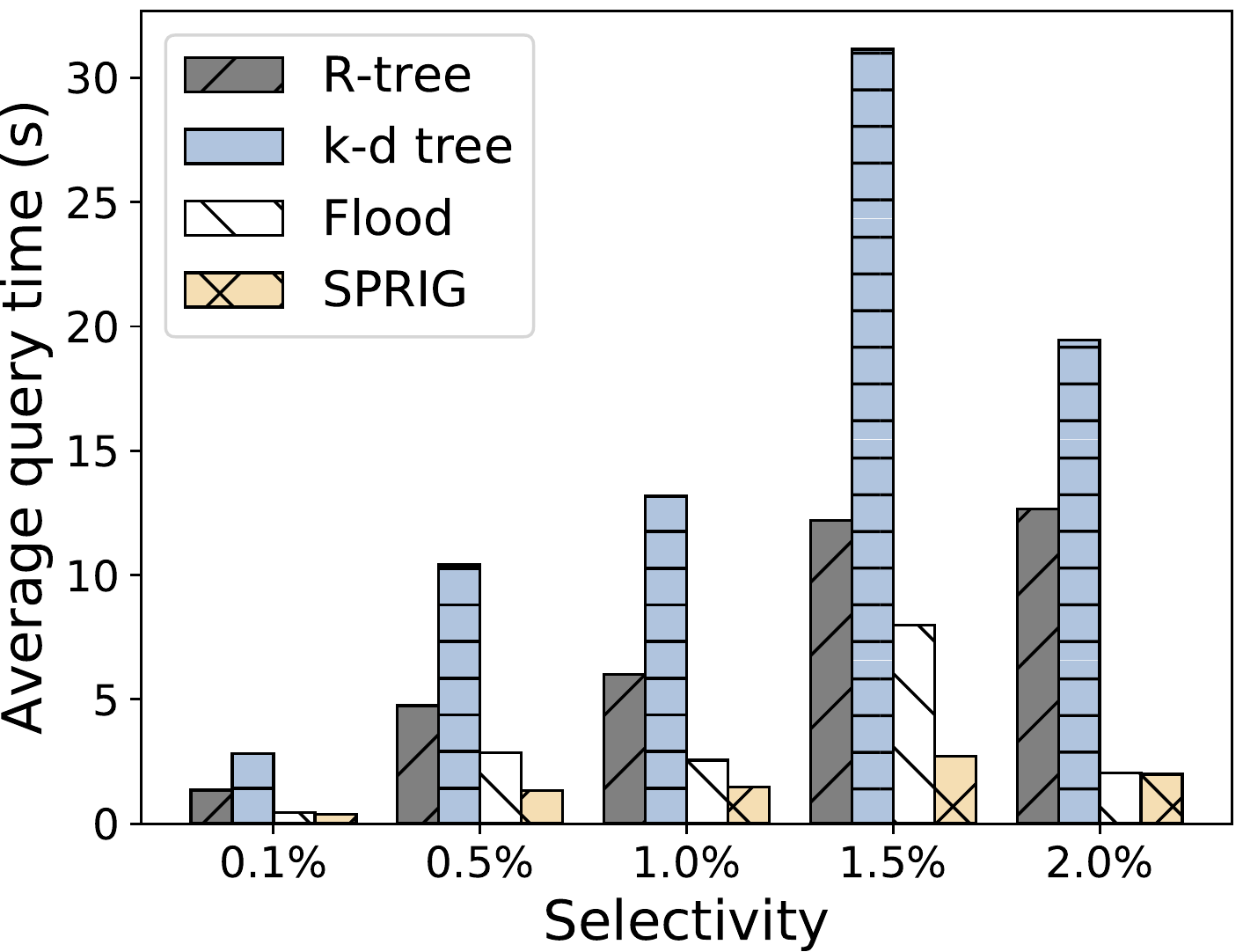}
    \label{fig.rq.20m}
    \end{minipage}%
}%
%\vspace*{-0.15cm}
\caption{Average query time of range query over different datasets}
\label{fig.eva.rq}
%\vspace{-0.2cm}
\end{figure*}

%Figures kNN query
\begin{figure*}[!t]
%\vspace{-0.4cm}
\centering
\setlength{\abovecaptionskip}{0.cm}
\subfigcapskip=-10pt
\subfigure[Tweet200K]{
    \begin{minipage}[t]{0.22\textwidth}
    \centering
    \includegraphics[height=0.76\textwidth,width=\textwidth]{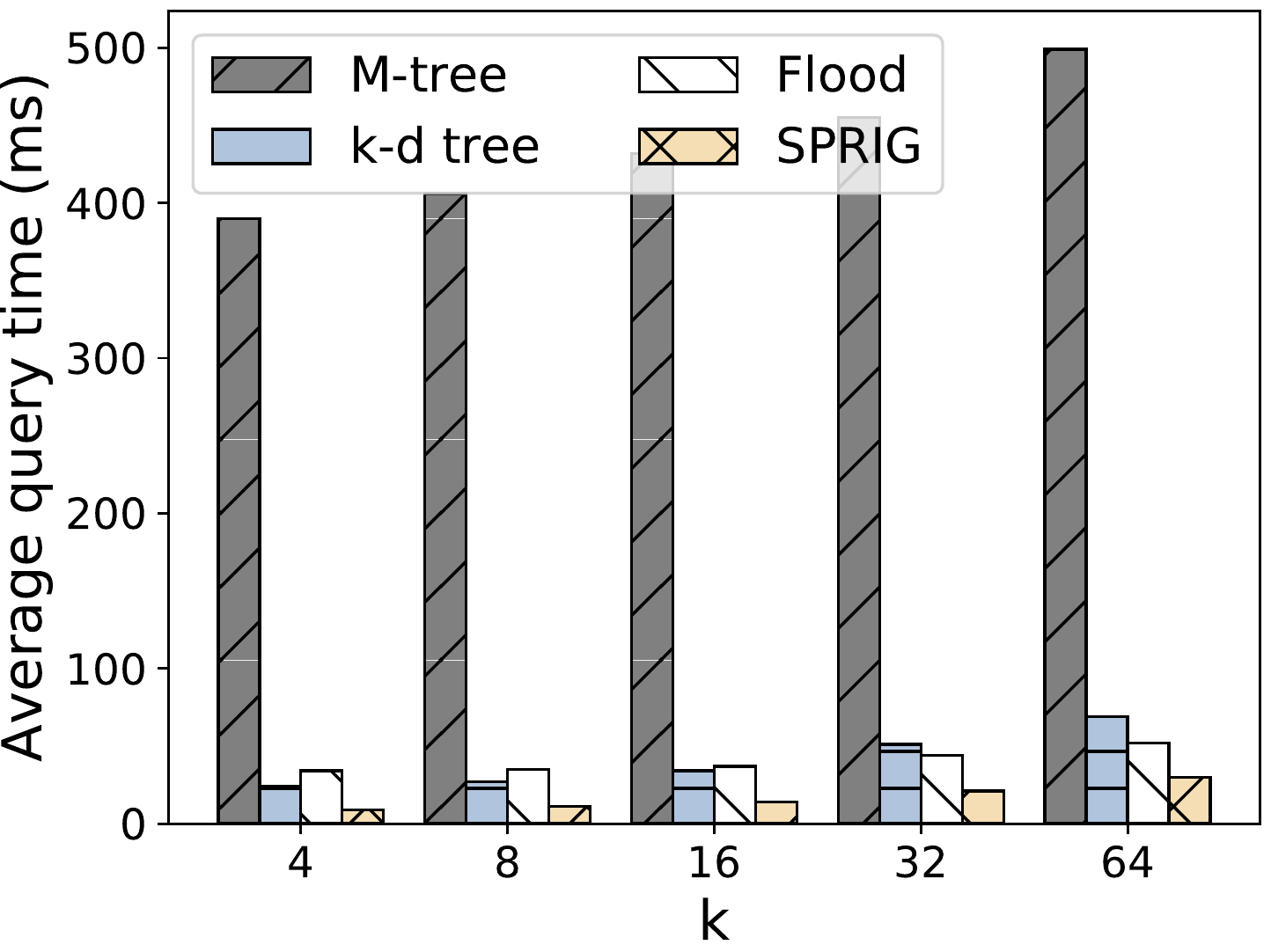}
    \label{fig.knn.200k}
    \end{minipage}%
}\hspace{10mm}%
\subfigure[Tweet2M]{
    \begin{minipage}[t]{0.22\textwidth}
    \centering
    \includegraphics[height=0.76\textwidth,width=\textwidth]{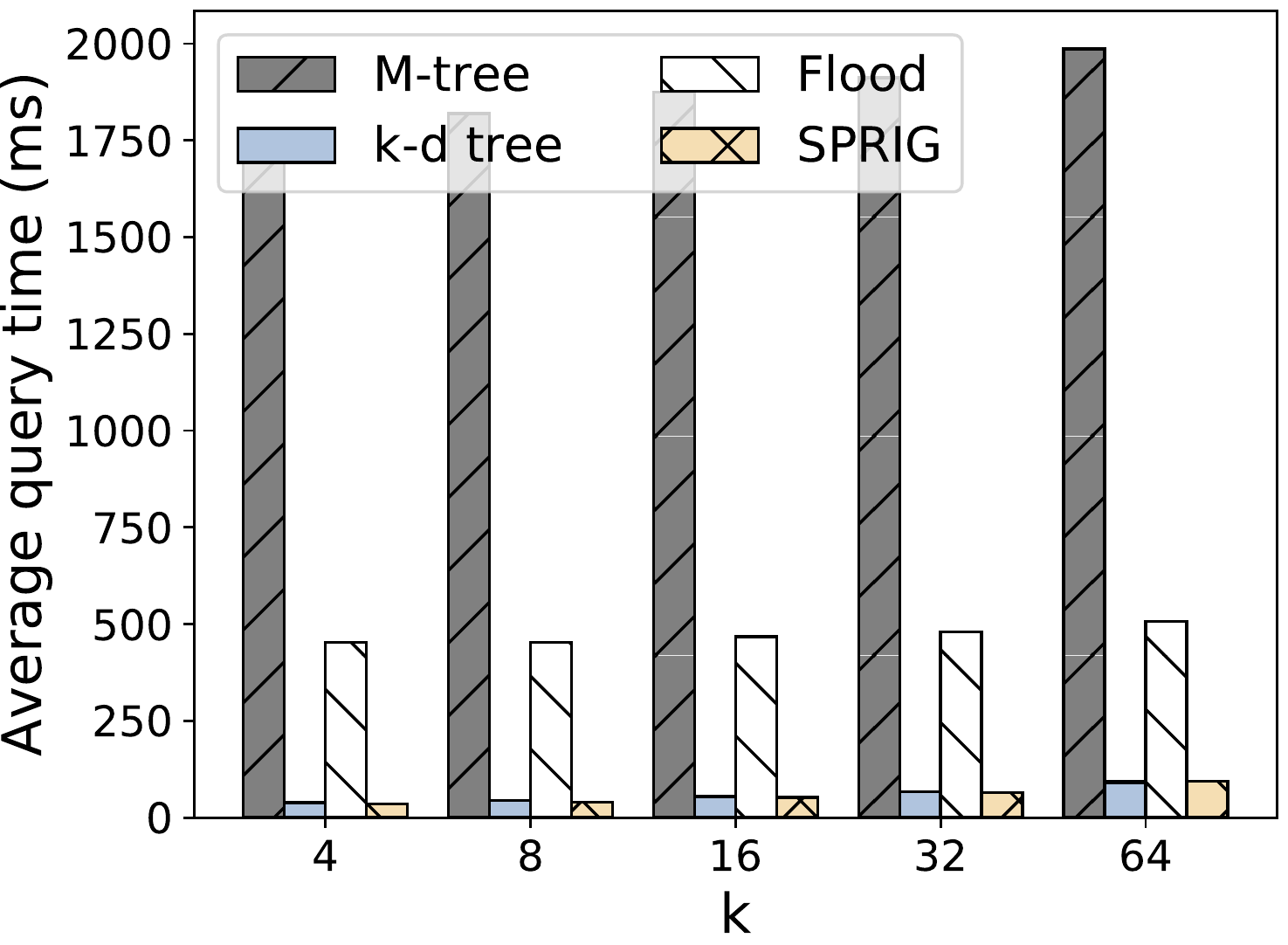}
    \label{fig.knn.2m}
    \end{minipage}%
}\hspace{10mm}%
\subfigure[Tweet20M]{
    \begin{minipage}[t]{0.22\textwidth}
    \centering
    \includegraphics[height=0.76\textwidth,width=\textwidth]{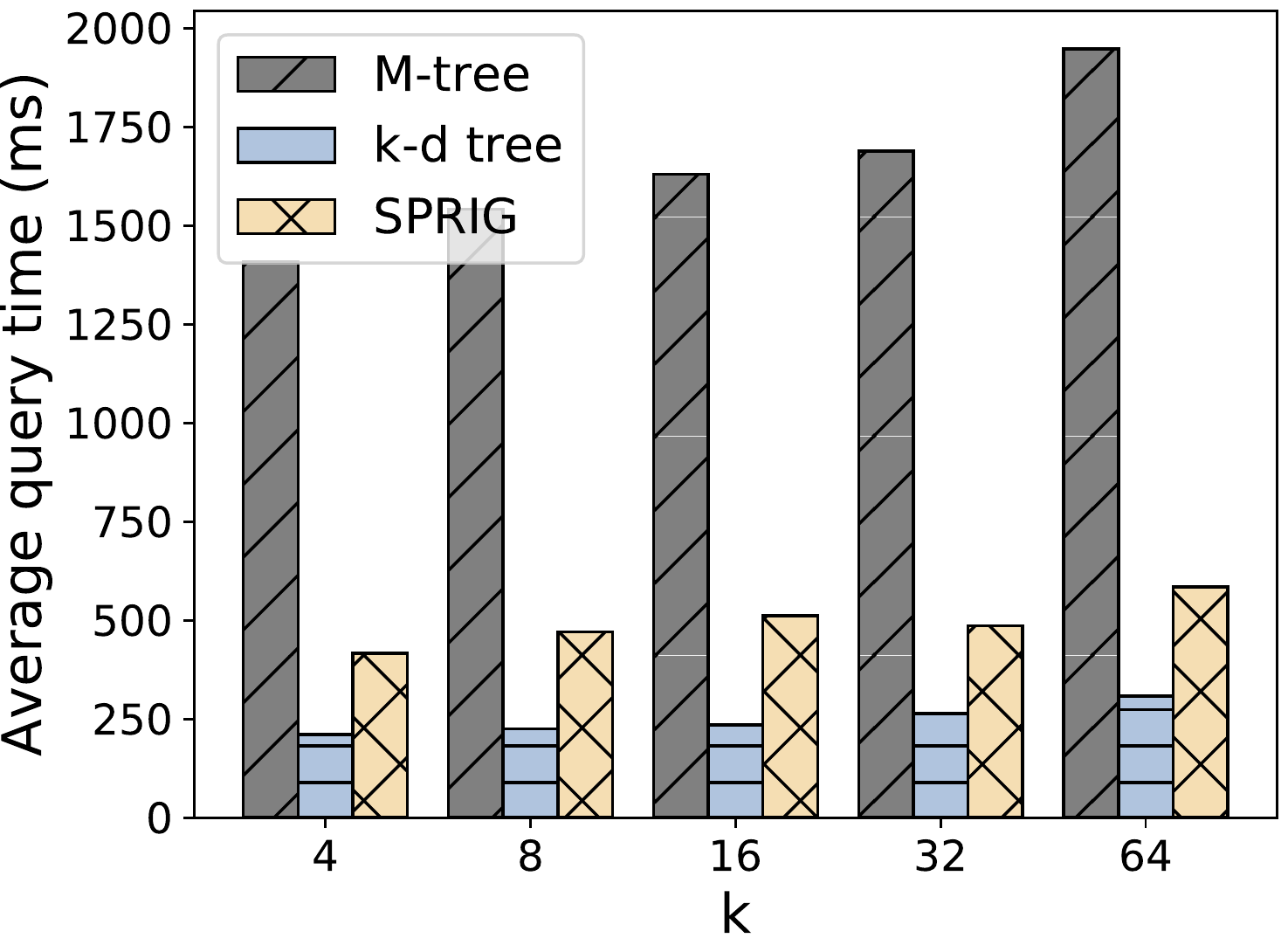}
    \label{fig.knn.20m}
    \end{minipage}%
}%
%\vspace*{-0.15cm}
\caption{Average query time of $k$NN query over different datasets. Note that, the average query time of Flood is individually shown in Table~\ref{tab.eva.flood} due to its significant execution time. }
\label{fig.eva.knn}
%\vspace{-0.2cm}
\end{figure*}

%Table 1
\begin{table*}[!t]
\centering
\vspace{-0.2cm}
\setlength{\abovecaptionskip}{0.1cm}
  \caption{Shepard and RBF Interpolation Functions}
  \label{tab.eva.sr}
  \begin{tabular}{ccccccc}
    \hline
    Metrics &Functions &$10 \times 10$&$20 \times 20$&$50 \times 50$& $100 \times 100$& $200 \times 200$\\
    \hline
   Execution time (ms) &Shepard & 6.7 & 25.8 &  161.2 &  646.3 & 2595.7 \\
   Execution time (ms) &RBF & 6.1 & 10.3 & 60.9 & 243.3 & Out of memory \\
  \hline
   Estimation Error &Shepard & 60 & 223 &  1322 & 5195 & 20587 \\
   Estimation Error &RBF & 1815 &6497& 30813& 116772 & Out of memory \\
  \hline
\end{tabular}
\vspace{-0.15cm}
\end{table*}

\subsection{Cost Model}
\label{sub.cost.model}
In this section, we build a cost model, which can be used to determine the number of columns in $x$ dimension and $y$ dimension, i.e., the values of $n$ and $m$.  Motivated by~\cite{NathanDAK20}, we model the execution time of performing a query over our index. Here, we take the range query workload as an example.  From Section~\ref{sub.index.query}, we know that the query time of range queries consists of four parts: 1) Prediction; 2) Local binary search; 3) Retrieve cells; 4) Scan and check data points in the intersected cells. Clearly, different $n$ and $m$ will generate different $\mathsf{B_x}, \mathsf{B_y}$. Consequently, the execution time of $\mathsf{F}_{in}$ and the local binary search will be affected. For ease of description, we denote the execution time of the spatial interpolation function as $\mathsf{T}(\mathsf{F}_{in}^{n \times m})$ and the execution time of local binary search as $\mathsf{T}(\mathsf{B}^{n \times m})$. Assume there are $N_i$ intersected cells and $N_c$ contained cells fall within the range query $\mathsf{Q}_r$. Meanwhile, we define  $N_p$ as the data points in the intersected cells.   Through simulations, we can obtain the average time of retrieving a cell and scanning a data point, which are denoted as $\mathsf{T}_r$ and $\mathsf{T}_s$, respectively.  Putting these four parts together, the time cost of performing a rang query is modeled as:
\[
	\textit{Time}  = \mathsf{T}(\mathsf{F}_{in}^{n \times m}) +\mathsf{T}(\mathsf{B}^{n \times m}) + \mathsf{T}_r \cdot (N_i + N_c) + \mathsf{T}_s \cdot N_p.
\]
Given a dataset $\mathcal{D}$ and a query workload $ \mathcal{W}$,  where $\mathsf{Q}_r \in \mathcal{W}$,  we expect to obtain the best layout parameter: $n \times m$ that makes $\textit{Time}$ have minimal average value.

\section{Evaluation}
\label{sec:eva}
In this section, we experimentally evaluate the performance of our index and compare it with alternative schemes in processing range and $k$NN queries.  Specifically,  we first explore the efficiency and accuracy of the typical spatial interpolation functions. Then, we compare SPRIG with traditional indices and Flood~\cite{NathanDAK20}, which is a recently proposed multi-dimensional learned index,  in terms of range and $k$NN query time and storage overheads.  We implemented all indexes in Java and evaluated them with in-memory versions.  To analyze the impact of dataset size on index performance, we adopt three Twitter datasets~\cite{twitter-dataset} consisting of tweets with their locations: \textit{Tweet200k}, \textit{Tweet2M}, and \textit{Tweet20M} that have 200k, 2M, and 20M spatial points, respectively. All experiments are conducted on a machine with 16 GB memory and 3.4 GHz Intel(R) Core(TM) i7-3770 processors and running Ubuntu 16.04 OS.

%Figures functions
\begin{figure}[!t]
\vspace{-0.1cm}
\centering
\setlength{\abovecaptionskip}{0.1cm}
\subfigcapskip=-10pt
\subfigure[]{
    \begin{minipage}[t]{0.22\textwidth}
    \centering
    \includegraphics[height=0.76\textwidth,width=\textwidth]{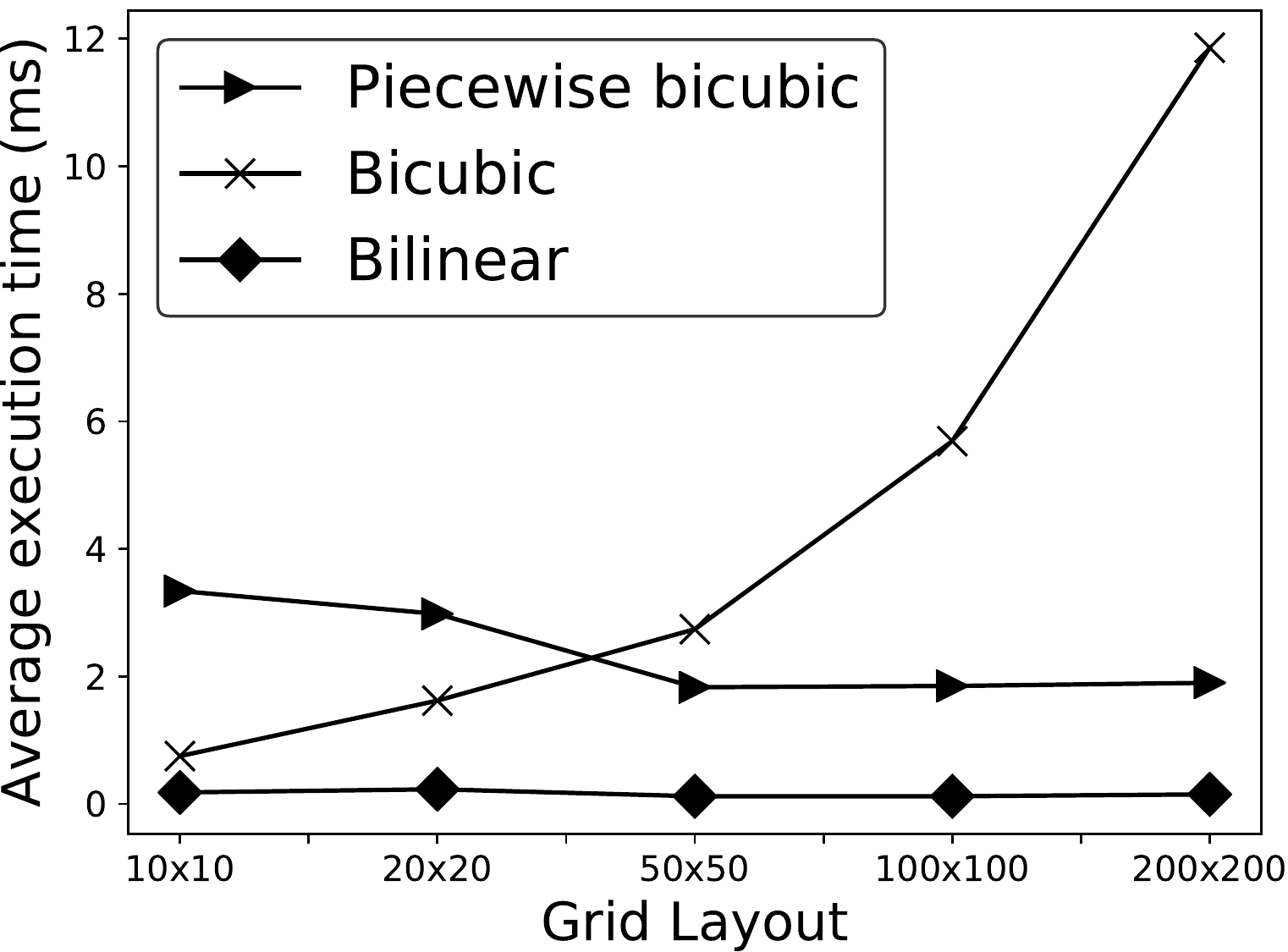}
    \label{fig.func.eff}
    \end{minipage}%
}%
\subfigure[]{
    \begin{minipage}[t]{0.22\textwidth}
    \centering
    \includegraphics[height=0.76\textwidth,width=\textwidth]{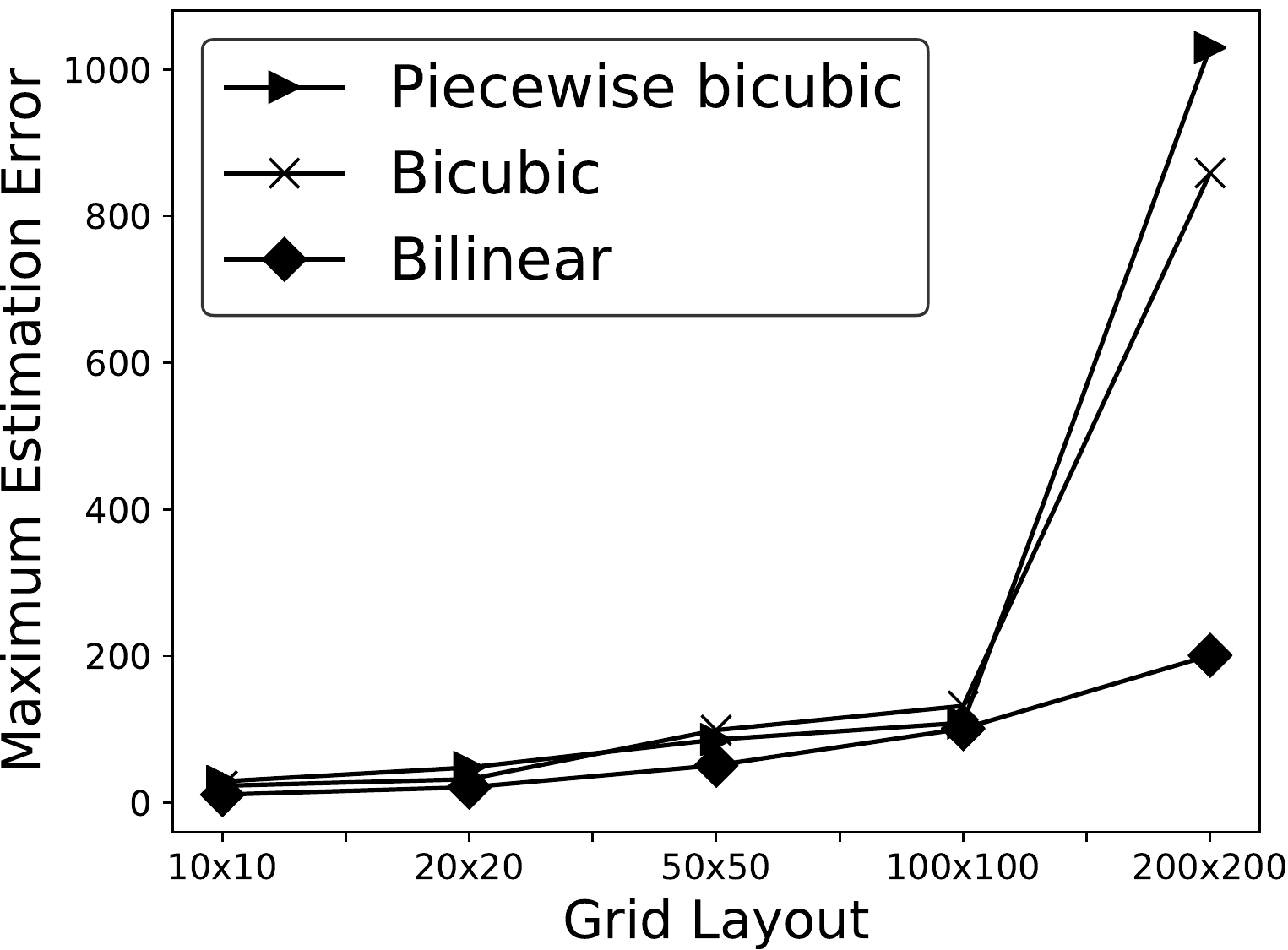}
    \label{fig.func.acc}
    \end{minipage}%
}%
\vspace*{-0.1cm}
\caption{Spatial Interpolation Functions. (a) Average Execution Time. (b) Maximum Estimation Error}
\label{fig.eva.func}
\end{figure}

\textbf{Efficiency and Accuracy of Spatial Interpolation Functions}. We evaluate five spatial (two-dimensional) interpolation functions on adaptive grids, i.e., \textit{bilinear interpolation},  \textit{bicubic interpolation}, \textit{piecewise bicubic interpolation}, \textit{Shepard interpolation},  and  \textit{radial based function (RBF) interpolation} ~\cite{Myers94}. Given a query workload, we expect to evaluate the average execution time and maximum estimation errors of these five functions varying grid layout from $10 \times 10$ to $200 \times 200$.  It is noted that, for ease of comparison, we adopt  $\textit{eg}$ (Eq.~\eqref{equ.eg}) instead of  $\textit{eg}_x$ and $\textit{eg}_y$ to evaluate the accuracy of  $\mathsf{F}_{in}$. All of these interpolation functions are evaluated on  \textit{Tweet200k}.   Fig.~\ref{fig.func.eff} shows the average execution time for three interpolation functions: \textit{bilinear}, \textit{bicubic}, and \textit{piecewise bicubic}, while Fig.~\ref{fig.func.acc} presents their accuracy. We exclude the \textit{Shepard} and \textit{RBF interpolation} functions, for which they have much larger execution time and estimation errors. However,  we still list their average execution time and estimation errors in Table~\ref{tab.eva.sr}.  From Fig.~\ref{fig.func.eff}, ~\ref{fig.func.acc},  and Table~\ref{tab.eva.sr}, we can see that the   \textit{bilinear interpolation} function has the best efficiency and accuracy in all grid layouts. Therefore, in the following comparisons, we apply  \textit{bilinear interpolation} function in our index. It is interesting that the maximum estimation error of \textit{bilinear interpolation} function is always $n$ + 1, which makes it quite suitable to reduce the execution time of the local search.

\textbf{Comparison on Range Query.}
For range queries, we compare our proposed index with other multi-dimensional indexes that support range queries over the  aforementioned three datasets. We choose R-tree, k-d tree, and Flood as the competing indexes, in which the first two are representative traditional indexes,  and the third one is a state-of-the-art learned index and similar to our work. In addition,  we consider five sets of queries with different selectivities \{0.1\%, 0.5\%, 1\%, 1.5\%, 2\%\}.  Aggregating these query sets as a big query workload,  we tune our index by the cost model (See details in Section~\ref{sub.cost.model}) to obtain the best grid layout. To be fair, we also train Flood by the approach proposed in~\cite{NathanDAK20}, which can obtain the best layout and error guarantee. Similarly,  the traditional indices,  R-tree and k-d tree, are tuned to obtain their best parameter, i.e., the maximum number of children for a node. Therefore, all indexes participating in the comparison are evaluated with their best parameters. From Fig.~\ref{fig.eva.rq}, we can see that SPRIG is always significantly faster than the traditional indexes in all selectivities, which is more advantageous on big data set. For example, on the \textit{Tweet20M}  dataset (Fig.~\ref{fig.rq.20m}),  our index achieves up to an order of magnitude better performance than k-d tree for range queries. Besides, our index outperforms Flood in most cases.  This benefit comes from the fact that Flood only learns the distribution of one dimension, while SPRIG learns the spatial distribution, which allows us to do more fine-grained filtering.

\textbf{Comparison on $k$NN Query.} For $k$NN queries, we replace R-tree with M-tree~\cite{CiacciaPZ97}. It is because R-tree is much slower than other indexes in $k$NN queries, and M-tree is a typical $k$NN-support index. Since Flood does not provide any detail about how to deal with $k$NN queries, we implement it with a similar $k$NN search strategy as our index. In the $k$NN query comparison, we consider  $k=\{4,8,16,32,64\}$.  Similar to the range query, we tune all the comparison indexes to obtain their best parameters. From Fig.~\ref{fig.eva.knn}, we can see that M-tree is more expensive than our index in executing $k$NN queries.  For Flood,  our index outperforms this learned index on all datasets. Since Flood is much slower than other indexes on the \textit{Tweet20M} dataset, we exclude it from Fig.~\ref{fig.knn.20m} and list its performance as follows.  
\begin{table}[!h]
\centering
\vspace{-0.15cm}
\setlength{\abovecaptionskip}{0.10cm}
  \caption{Average query time of Flood on \textit{Tweet20M}}
  \label{tab.eva.flood}
  \begin{tabular}{cccccc}
   \hline
    k  & 4 & 8 & 16 & 32 &64 \\
    \hline
    Query Time (ms) &4590 & 4796 & 4818 &  4778 & 5073 \\
   \hline
\end{tabular}
\vspace{-0.15cm}
\end{table}
As shown in Fig.~\ref{fig.eva.knn} and Table~\ref{tab.eva.flood}, our index is around 2$\times$, 5$\times$, and 9$\times$ faster than Flood on the \textit{Tweet200k},  \textit{Tweet2M}, and  \textit{Tweet20M} dataset, respectively. For k-d tree,  our index is at least 2$\times$ faster on the \textit{Tweet200k} dataset and is slightly faster on the \textit{Tweet2M} dataset. For the \textit{Tweet20M} dataset, although our index still has a huge advantage over M-tree and Flood in the query performance, it is not as good as k-d tree. However, k-d tree has a significantly higher storage footprint to achieve such query performance.   We demonstrate the storage overheads of these indexes in the next section.

\textbf{Storage Overhead.} Due to the limited space, we only compare the storage overheads of different indexes on the \textit{Tweet20M} dataset, which actually has a similar relationship between indexes on the other two datasets. In the range and $k$NN query comparisons,  we use the corresponding query workload to tune these indexes for their best parameters. For example,  our index has the best grid layout $710 \times 690$ on rang query workload, while it is $3000 \times 10$ for $k$NN query workload. Since different parameters of one index may lead to different storage overheads,  we compare the storage overheads of these indexes on the range and $k$NN query workloads, separately, as shown in Table~\ref{tab.store.rq} and Table~\ref{tab.store.knn}, respectively.  To obtain the storage footprint of a tree index, we evaluate the necessary storage of one node, e.g., the minimum bounding rectangle (MBR) of R-tree,  and count the total number of internal nodes by traversing the tree. For Flood, it has two components: a  \textit{FITing-tree} on one dimension and a table to map a cell to the covered data records. For our index,  there are three components: a grid layout $\mathsf{G}_{n \times m}$, a table $\mathcal{T}$ for managing the cells, and a spatial interpolation function  $\mathsf{F}_{in}$.  The storage consumption of these two learned indexes is related to their grid layout.  Table~\ref{tab.store.rq} and Table~\ref{tab.store.knn} show that both the learned indexes have less storage overheads than traditional indexes on the range and $k$NN query workloads. Although our index consumes more storage compared to Flood,  we achieve better query performance with an acceptable storage overhead. Recall that in the $k$NN query time comparison, our index is slower than k-d tree on  \textit{Tweet20M}. However, as shown in Table~\ref{tab.store.knn}, the storage overhead of k-d tree is three orders of magnitude larger than that of our index, which renders it challenging to use in practice. Thus,  our index is more practical than k-d tree with big datasets.

\begin{table}[!t]
\centering
\setlength{\abovecaptionskip}{0.1cm}
  \caption{Storage of indexes on range query workload}
  \label{tab.store.rq}
  \begin{tabular}{ccccc}
    \hline
    Index  & R-tree & k-d tree & Flood & SPRIG  \\
   \hline
    Storage Overhead (MB) & 7.95 & 305.17 & 0.11 &  6.30  \\
  \hline
\end{tabular}
\end{table}

\begin{table}[!t]
\setlength{\abovecaptionskip}{0.1cm}
\centering
  \caption{Storage of indexes on $k$NN query workload}
  \label{tab.store.knn}
  \scalebox{1.0}{
  \begin{tabular}{ccccc}
     \hline
    Index  & M-tree & k-d tree & Flood & SPRIG  \\
   \hline
    Storage Overhead (MB) & 65.89 & 305.17 & 0.05 &  0.60  \\
  \hline
\end{tabular}}
\end{table}

\section{Conclusion}
\label{sec:con}
In this paper, we have proposed a new learned model that can learn the spatial distribution of the spatial data directly. Based on the learned model, we have built a novel learned spatial index SPRIG and designed the range and $k$NN query execution strategies over the index. Our experimental results suggest that 1) the \textit{bilinear interpolation} function is the best option as the spatial learned model compared with the other spatial interpolation functions; 2) our index SPRIG is efficient with the relatively small storage footprint.  In our future work, we expect to further reduce the average query time of $k$NN queries on big datasets and make our index more flexible.

%\bibliographystyle{IEEEtran}
%\bibliography{IEEEabrv,interpolation}


%\nocite{*}

%%%%%%%%%%%%%%%%%%%%%%%%%%%%%%%%%%%%%%%%%%%%%%%%%%%%%%%%%%%%%%%%%%%%%%%%%%%%%%%%
\end{document}